\documentclass{article}

\usepackage{arxiv}

\usepackage{amsmath}		   

\usepackage[utf8]{inputenc} 
\usepackage[T1]{fontenc}    
\usepackage{hyperref}       
\usepackage{url}            
\usepackage{booktabs}       
\usepackage{amsfonts}       
\usepackage{nicefrac}       
\usepackage{microtype}      
\usepackage{cleveref}       
\usepackage{lipsum}         
\usepackage{graphicx}
\usepackage{natbib}
\usepackage{doi}
\usepackage{float}

\usepackage{xcolor}

\title{Protecting Small Organizations from AI Bots with Logrip: Hierarchical IP Hashing}

\date{August 4, 2025}

\newif\ifuniqueAffiliation

\uniqueAffiliationtrue

\ifuniqueAffiliation 

\author{ \href{https://orcid.org/0000-0002-0449-981X}{\includegraphics[scale=0.06]{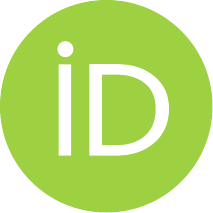}\hspace{1mm}Rama Carl Hoetzlein} \\
	Quanta Sciences\\
	Ithaca, NY 14850 \\
	\texttt{ramahoetzlein@gmail.com} \\
}
	
\else

\fi

\renewcommand{\shorttitle}{Protecting Small Organizations from AI Bots}

\hypersetup{
pdftitle={Protecting Small Organizations from AI Bots with Logrip: Hierarchical IP Hashing},
pdfsubject={q-bio.QM, q-bio.NC},
pdfauthor={Rama Carl Hoetzlein},
pdfkeywords={security, defense, web server, crawlers, AI bot, rate limiting, human-computer interaction, HCI, access logs, web analytics, data visualization},
}

\begin{document}
\maketitle

\begin{abstract}
Small organizations, start ups, and self-hosted servers face increasing strain from automated web crawlers and AI bots, whose online presence has increased dramatically in the past few years. Modern bots evade traditional throttling and can degrade server performance through sheer volume even when they are well-behaved. We introduce a novel security approach that leverages data visualization and hierachical IP hashing to analyze server event logs, distinguishing human users from automated entities based on access patterns. By aggregating IP activity   across subnet classes and applying statistical measures, our method detects coordinated bot activity and distributed crawling attacks that conventional tools fail to identify. Using a real world example we estimate that 80–95\% of traffic originates from AI crawlers, underscoring the need for improved filtering mechanisms. Our approach enables small organizations to regulate automated traffic effectively, preserving public access while mitigating performance degradation.
\end{abstract}


\section{Introduction}
With the proliferation of automated web crawlers, AI bots, and malicious scanning activities, small organizations and self-hosted servers are increasingly vulnerable to overwhelming traffic loads.  Recent studies have found that, for the first time, more than 51\% of Internet traffic can be attributed to non-human activity (Imperva 2025, Bad Bot report), and nearly 37\% can be described as "bad bot" behavior.

Traditional security approaches such as throttling fail to adequately capture and mitigate these threats. Our method introduces a novel approach to web server security by focusing on the distinction between human versus machine access patterns. We leverage data visualizations to analyze server event logs and filter IPs and subnets with mechanical access patterns. With a specific case study we demonstrate traffic reductions of 94\%, significantly greater than rate limiting alone.

Standard web statistics tools, such as GoAccess, focus on general traffic analysis but lack the granularity to differentiate AI bots that respect throttling limits yet still degrade performance through sheer volume. Our approach groups access patterns by IP and aggregates them hierarchically across subnet classes, applying statistical measures at each level. This enables detection of coordinated bot activity and distributed crawling attacks, which often evade conventional detection mechanisms. Through visual analysis, we distinguish between human and non-human activity, empowering organizations to regulate and filter harmful traffic effectively.

\section{Background}

Prior research has extensively characterized automated browsing behavior and its implications for web security. \cite{Li2021} explores the nature of good and bad bots, demonstrating that a significant percentage, up to 87\%, of malicious IPs are not listed on common, publicly available blocklists. \cite{Owen2022} surveys botnets, categorizing them by activity and method of access. Our approach builds on this by directly analyzing raw access patterns to generate useful blacklists.

\subsection{Server log analysis}

There is significant prior research on extracting meaningful patterns from web server access logs. \cite{Nguyen2018} employs a statistical approach, while \cite{Seyyar2018} highlight the limitations of existing methods in detecting frequency-based attack patterns. \cite{Meyer2008} introduces anomaly detection using dynamic rules, requiring a non-attack baseline, which is not always feasible. \cite{Yen2013} leverage large-scale log analysis, but their methods are tailored for enterprise environments rather than small-scale servers. \cite{Du2017} utilizes deep learning to build blacklists, although its pre-filtering struggles to distinguish legitimate traffic from attacks. We observe that the difficulty in generating training data for AI automated blocking resides in the challenge of detecting whether the actions of a given IP can be classified as mechanical or human.

\subsection{Web crawling}

In the domain of web crawling, \cite{Wolf2002} and \cite{Ro2018} describe strategies for developing distributed crawlers, while \cite{Kausar2013} provide a broad review of crawling techniques. \cite{Gold2017} raises ethical and legal concerns surrounding web crawling but acknowledges that, regardless of intent, excessive crawler traffic can cripple small servers. \cite{Chang2015} discovered that IP addresses and families are heavily reused by bot nets through the analyzing of bots in the wild, a finding confirmed in our case study. We address this issue specifically by developing methods to identify subnet family data center attacks. In the work of \cite{Kim2018}, \cite{Chang2015} and \cite{Stange2014}, visualizations of traffic by time and host IP reveal insights not captured by standard web statistics, motivating our human-computer vision approach to web traffic analysis.

\subsection{Rate limiting}

Early development of the Internet focused on limiting network and server loads to maintain operability, with the introduction of the leaky bucket algorithm \cite{Turner1986}. A series of "congestion collapses" in the 1980s required evolution of the TCP/IP stack to handle high speed traffic \cite{Jacobson1988}. Later work began to understand the data flow limitations of web servers \cite{Banga1997}, and methods for controlling overflow \cite{Iyer2001}. Rate limiting has become common practice for throttling web server traffic, yet we demonstrate that rate limiting is no longer able to mitigate the volume of modern automated crawler activity.

\begin{figure*}[!h]
  \includegraphics[width=\textwidth]{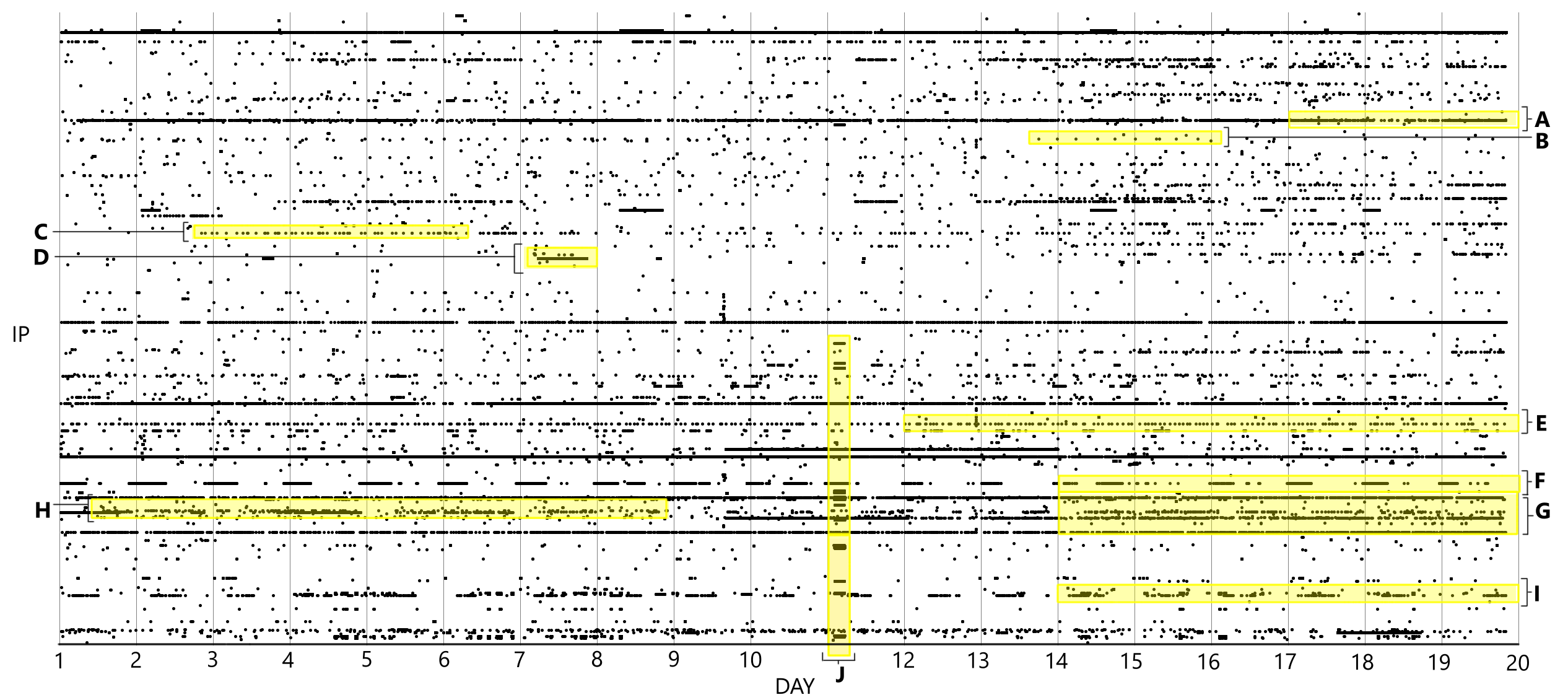}  
  \caption{Data visualization of raw access logs plotted with time in days (x-axis) versus IP address (y-axis) reveal consistent mechanical patterns attributable to AI crawlers and bot nets. Each highlighted, marked region shows a distinct yet recognizable mechanical access pattern inconsistent with human behavior. The majority of traffic, estimated at 80-95\%, for this 20 day period are likely AI crawlers and bots.}
  \label{fig_overview}

  \begin{center}
  \label{tab:regions}
  \begin{tabular}{lllll}
  \toprule
  Marked Region & Daily frequency & Multi-day & Daily range & IP usage \\
\midrule
A & rapid, repetitive & consecutive & 24 hr/day & single \\
B & infrequent, regular & consecutive & 24 hr/day & single \\
C & scattered, consistent & consecutive & 24 hr/day & single \\
D & rapid, repetitive & single day & >10 hr/day & single \\
E & scattered, frequent & consecutive & 20-24 hr/day & single \\
F & rapid, repetitive & consecutive & 3 hr/day, shifting & single \\
G & rapid, repetitive & consecutive & 24 hr/day & multiple IPs, narrow \\
H & rapid, repetitive & consecutive & >18 hr/day & multiple IPs, wide \\
I & rapid, repetitive & consecutive & 8-12 hr/day & multiple IPs, wide \\
J & short term & single day & <4 hr/day & full IP range \\
  \bottomrule
  \end{tabular}
  \end{center}
  \caption{Types of access patterns by marked region in Figure 1. Daily frequency refers to access rate and pattern within a given day. Multi-day refers to behavior over many days. Daily range refers to the portion of the day in which activity is observed. IP usage is the number of distinct IP addresses used in an attack.}
\end{figure*}

\section{Motivation}

This work is motivated by a real world case study of access logs from a public, community-based organization, The Community Science Institute of Ithaca, New York, an organization that provides public water quality analysis and programs that advocate for citizen science and education. Despite the modest scale of the Community Science Institute (CSI) we observed a disproportionately high traffic volume, with over 150,000 page hits over a period of 20 days. This corresponds to an average of roughly 7500 pages per day, whose origin covers the entire globe. Since the Community Science Institute provides regional water quality data for New York State the global scale of the traffic is easily attributable to AI crawlers rather than human users. We estimate that 80-95\% of observed traffic derives from AI bots and crawlers.

The Community Science Institute (CSI) provides public, high quality, laboratory validated data for water quality analysis. This data is human curated and takes time to produce thus making it a valuable resource for AI training data. The data created by CSI is in the public domain with general access permitted and encouraged; tools for bulk download are even provided. Despite this open data policy the sheer volume of traffic from AI crawlers was found to degrade the server performance for real human users. Thus the primary policy motivations for blocking AI bot traffic at CSI are related to usability rather than data protection. 

Preliminary analysis of server logs through data visualization, Figure 1, reveal that AI crawler activity is predominantly mechanical and repetitive, characterized by regular access intervals that exceed human browsing patterns. Conventional throttling mechanisms fail to mitigate these issues, as distributed bots remain within permitted limits while collectively generating unsustainable loads. Moreover, existing statistical tools, such as GoAccess, lack the ability to differentiate between human and automated traffic, rendering them ineffective for targeted filtering.

A detailed visual analysis of the source logs reveals activity that is mechanical, but with different signatures identified in Table 1. This led us to reconsider the blocking problem as the inverse of detecting human activity. That is all mechanical patterns are revealed as either too fast, making use of consecutive days, many hours per day, ignoring day/night cycles, or employing many IPs simultaneously; actions which human users tend to avoid.

\section{Algorithm Design}

This work focuses on the problem of detecting mechanical access patterns in requests to web servers regardless of whether the origin is benign or malicious. We ignore the nature of the requests and consider only the pattern of access. For this work, we consider crawlers and bots as isolated or distributed machines whose primary purpose is to collect data - without knowing whether the origin is agent-guided or whether the intended purpose is for search engines, databases or AI training. Our goal is simply to distinguish human from mechanical access patterns so that the latter may be blocked according to a given policy.

\subsection{Behavioral metrics}

Metrics are developed to detect machine access patterns in contrast to human patterns of computer use, for which the field of Human-Computer Interaction (HCI) can provide valuable data. Beauvisage et al. examines computer use in daily life, finding an average use of 3 hours 30 minutes per day in 2009 \cite{Beauvisage2009}. Like all human activity computer use is connected to the circadian rhythm which regulates biological sleep cycles, limiting the human use of computers at night. Studies show that violating this rhythm disturbs sleep quality \cite{Mesquita2007}. While these measures are statistical, not rigorous, they can be used together to establish bounds for non-human activity. A set of policies $\textbf{P}_{policy}$ are implemented using metric parameters, $P_{metric}$, that attempt to capture the distinction between human and machine access patterns. Four policies are developed:

\textit{Smart throttling}. Throttling assumes that machines make requests significantly faster than humans, as measured from a specific IP. However, we noticed that the majority of requests in Figure 1 were \textit{observing} rate limits. Presumably their designers are aware that ignoring rate limits will result in blocking and therefore follow other tactics. We suggest smart throttling, which limits by rate \textit{if} a given IP exceeds a daily average hit threshold.

(1) \hspace{0.5cm} $\textbf{P}_{\text{smart throttle}}$ = IF: ave hits per day > $P_{\text{max daily ave}}$ AND hits per min > $P_{\text{max hits per min}}$

\textit{Daily total}. Another common metric employed by web servers assumes that malicious machines are able to make many more total daily requests than humans. This may fail to account for well-behaved machines that act human by mimicry with regular, low frequency requests. Nonetheless we include this metric to block especially aggressive machines.

(2) \hspace{0.5cm} $\textbf{P}_{\text{daily total}}$ = IF: daily total hits > $P_{\text{max daily total}}$

\textit{Daily range}. Based on the observation that humans use computers for a limited duration of the day, we introduce a metric to capture continual daily activity \cite{Beauvisage2009}. Machines are able to sustain long duration, continual access patterns whereas humans do not. The daily range computes the span from the earliest to latest daily visit. We account for midnight rollover by also subtracting the largest daily gap.

(3) \hspace{0.5cm} $\textbf{P}_{\text{daily range}}$ = IF: (last daily hit time - first daily hit time) - largest gap > $P_{\text{max daily minutes per day}}$

\textit{Consecutive days}. Machines are observed to make requests over more than five continual days, whereas humans usually break this pattern due to weekends. Since this is not strictly true we combine this with a daily range that well exceeds human capacity. This pattern encodes the observation that many crawlers or data centers operate 24/7 (twenty-four hours a day, seven days a week).

(4) \hspace{0.5cm} $\textbf{P}_{\text{consecutive}}$ = IF: consecutive days > $P_{\text{max consec days}}$ AND consecutive daily range > $P_{\text{max consec minutes per day}}$

\begin{figure}
  \includegraphics[width=0.95\textwidth]{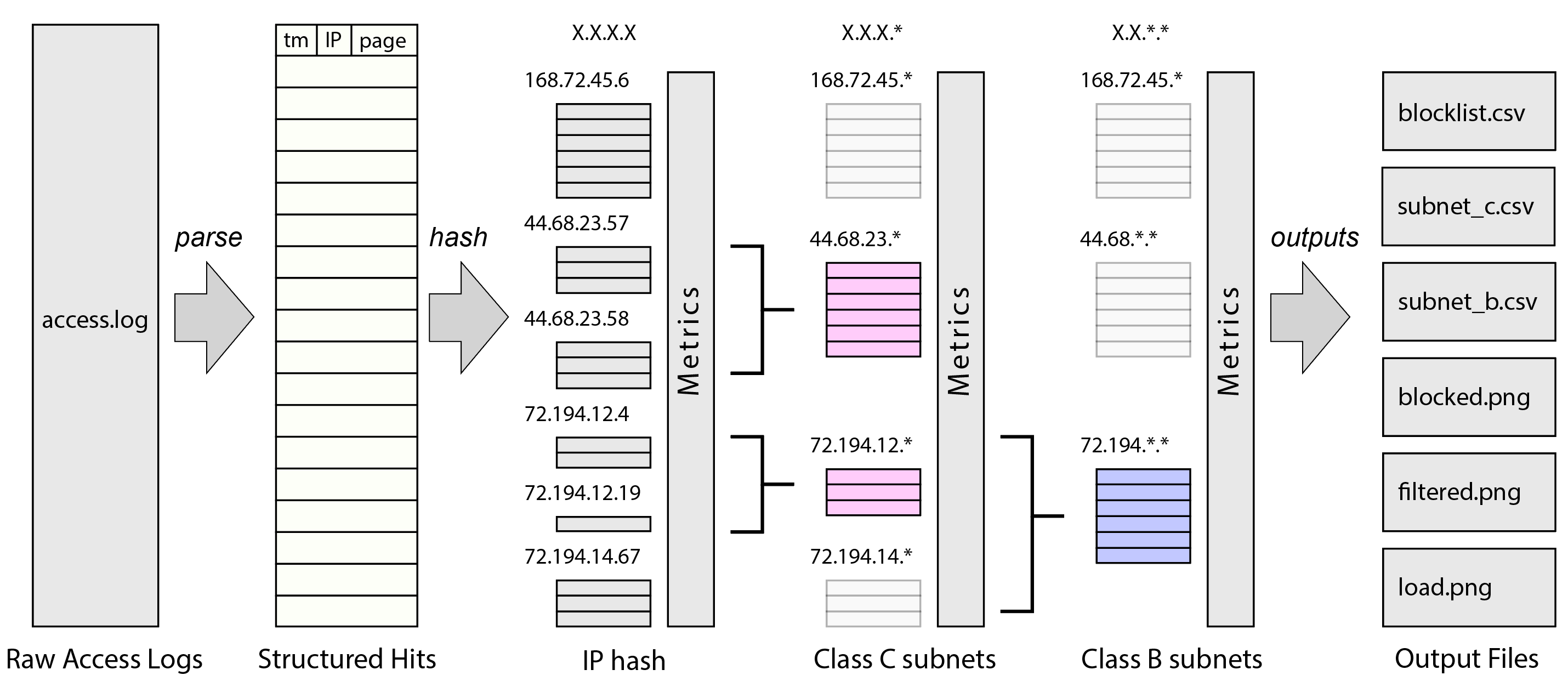}  
  \caption{Hierarchical IP Hashing of addresses by subnet. Raw access logs are parsed into page hits by time and IP (left). The first level hash organizes hits by IP address while retaining a sorted, temporal list of pages for each IP. Frequency and statistical metrics are applied after each hashing stage using a \textit{scoring algorithm}. Recursively, the next level hash merges Class C subnets while also aggregating and retaining pages at that level. The process of hash-sort-metric-scoring proceeds at each level up to Class B subnets. The output is a set of files containing the final blocklist, subnet lists (for further analysis), data visualization plots of blocked and filtered outcomes, and graphs of estimated server load before and after.}
  \label{fig_method}
\end{figure}

\subsection{Sorting, Binning \& Scoring Algorithm}
A scoring algorithm takes the structured visits derived from access logs and produces a set of metrics and blocking scores for each IP. Our goal is to score based on daily metrics, therefore we hash page hits and sort them by date and time. The steps are:

\begin{enumerate}
\item Hashing - All visits are hashed by IP address into a key-value map (Figure 2, IP Hash), while associating a list of page names and times with each unique IP.

\item Sorting \& Binning - The page list for each IP is sorted by time and binned by day. This prepares the list for computing daily policy metrics.

\item Metrics - The policy metrics are computed over each day for a given IP address. When a single IP covers multiple days, these metrics are computed separately per day and the daily minimum and maximum of each metric is recorded with the IP.

\item Scoring - A decision to block-or-accept is made for each IP if \textit{any} of the $\textbf{P}_{policy}$ thresholds was exceeded. The metric parameters $P_{metric}$ are input to determine how aggressive or lenient the scoring will be.
\end{enumerate}

The scoring algorithm operates on a list of time-stamped page names for each IP.  The output of scoring, stored in the same hash table, are the metric scores and the block-or-accept decision.

\pagebreak
\subsection{Hierarchical IP Hashing}
\label{hierarchical}

After initial experiments, it was observed that the new hashing and policy scoring performed better than simple throttling, yet a significant amount of traffic could still be visually identified as mechanical that was not blocked. This corresponded to sections F, G and H in Figure 1. These regions can be identified as data centers having the similar mechnical patterns yet utilizing multiple IPs.

We extend our core approach with Hierarchical IP Hashing in order to detect distributed bot activity. Data centers are known to circumvent single IP limits by employing many machines in tandem \cite{Chang2015}. We therefore hash IPs within subnets and perform metrics on that aggregated daily activity to capture patterns from multiple IPs.

After initial IP hashing and scoring, we progressively mask the least significant bytes of IP addresses to group access records at class C, B, and A subnet levels, as shown in Figure 2. We then progressively hash and score each level over the aggregated page/visit data for IP groups. In this way one can identify coordinated crawler behavior that is imperceptible at the individual IP level. The scoring algorithm above, including hasing, sorting, metrics and scoring, is applied at each subnet level. Class A subnets are hashed for recursive completeness, but we do not compute metrics or block at this level since the number of IPs is too broad.

\section{Implementation}

We introduce Logrip, a tool for analyzing, identifying, and blocking IP traffic based on behavioral analysis. Our technique combines several  metrics into a unified score which is applied to each IP. We extend this with hierarchical IP hashing, a method to aggregate and apply the same scoring technique to coordinated data center attacks from multiple IPs within Class C and Class B subnets. We also present novel comparative data visualizations, integrated with our tools, to provide a visual summary of reductions in mechanical access patterns.

Logrip reads and parses raw access logs produced by web servers. A dynamic regular expression parser handles different log formats. Access logs are parsed for each visit date, time, source IP, and page name and then placed into a large array of structured hits and sorted by time, as shown in Figure 2. A format string is provided in a config file to match a specific log format. Capture groups within the format string parse out specific log information, using the pre-defined groups given in Table 1. This allows for flexible handling of arbitrary log formats.

\begin{table}[h!]
\centering
\begin{tabular}{|l|l|}
\hline
\textbf{Capture group} & \textbf{Description} \\
\hline
\{X.X.X.X\} & IP address \\
\{AAA\} & Name of user/client \\
\{GET\} & Type of HTTP request. Will match to GET, POST or HEAD \\
\{PAGE\} & Page retrieved \\
\{PLATFORM\} & Platform details \\
\{DD/MMM/YYYY\} & Date in 12/Jan/2025 format \\
\{YYYY-MM-DD\} & Date in 2025-01-12 format \\
\{HH:MM:SS\} & Time in 24-hour HH:MM:SS format \\
\{RETURN\} & HTTP return code \\
\{BYTES\} & Number of bytes in return \\
\{NNN\} & Any captured number \\
* & Match any number of characters. Not captured. \\
\texttt{literal} & Match specific words or numbers. Not captured. \\
\hline
\end{tabular}
\caption{Capture groups for dynamic parsing.}
\end{table}

Policy parameters are set empirically to determine the lenience or aggressiveness of the blocking actions. The policy parameters, units and default values are provided in Table 2. For all parameters lower values result in more aggressive blocking. 

\begin{table}[h!]
\centering
\begin{tabular}{|l|l|l|l|}
\hline
\textbf{Policy parameter} & \textbf{Unit} & \textbf{Policy type} & \textbf{Default Value} \\
\hline
min ip b & \# machines & B-subnet filter & 1024 \\
min ip c & \# machines & C-subnet filter & 3 \\
max ip c & \# machines & C-subnet filter & 80 \\
max robot & \# hits per IP & robots filter & 10 \\
max daily & \# hits per day & daily hits & 100 \\
max daily range & minutes per day & range check & 360 \\
max consec days & days & consecutive & 5 \\
max consec range & minutes per day & consecutive & 240 \\
max daily ave & \# ave hits per day & smart throttling & 40 \\
max daily ppm & \# daily hits per min & smart throttling & 40 \\
\hline
\end{tabular}
\caption{Policy parameters with default values}
\end{table}

\subsection{Visualizations}

Scatter plots of time versus source IP address, found in Kim 2018, provided inspiration for data visualization as a way to identify mechanical access patterns through human visual recognition. Humans are able to easily recognize repetitive mechanical patterns in noise - appearing as horizontal, dotted lines in such plots - and providing a way to empirically study the effectiveness of different policy settings.
These visualizations represent the entire request log without statistical oversimplification. Color coding the blocking actions that were taken makes it easy to identify which patterns have been removed.

\subsection{Output products}

Logrip outputs several products to summarize actions taken. To develop a useful public tool all outputs are generated as either txt, csv, or png files for data visualizations. The outputs produced are:

\begin{itemize}
\item Blocklist - A list of IP addresses, with /16 and /24 masking, for blocklists generated at the individual, class B and class C levels. To keep the list short, class B/C subnets override individual IPs, which are then reported if they are blocked and unique.

\item Subnet B/C lists - A list of IP subnet addresses, which includes the number of machines in the group, and metric scores. For class B, we also output the sub-IPs and page numbers which are useful for reverse domain name lookup to identify specific organizations managing each subnet.

\item Blocked data vis - A scatter plot, in png format, of the blocked IP addresses. Color coding of red for single IPs, purple for class C subnets, and blue for class B subnets, this visualizes the total blocking actions were taking.

\item Filtered data vis - A scatter plot, in png format, of the allowed IP addresses and their visits. Ideally, this plot will reveal that the majority of mechanical, dense, repetitive patterns have been removed and the remaining data points are sporadic and infrequent; likely to be human.

\item Loading graph - A graph, in png format, of the estimated load on the server over time, with and without the blocklist. The loading at each moment in time t is estimated as the sum of currently active visits taking a fixed duration of server time to respond to. 
\end{itemize}

\begin{figure}
  \includegraphics[width=\textwidth]{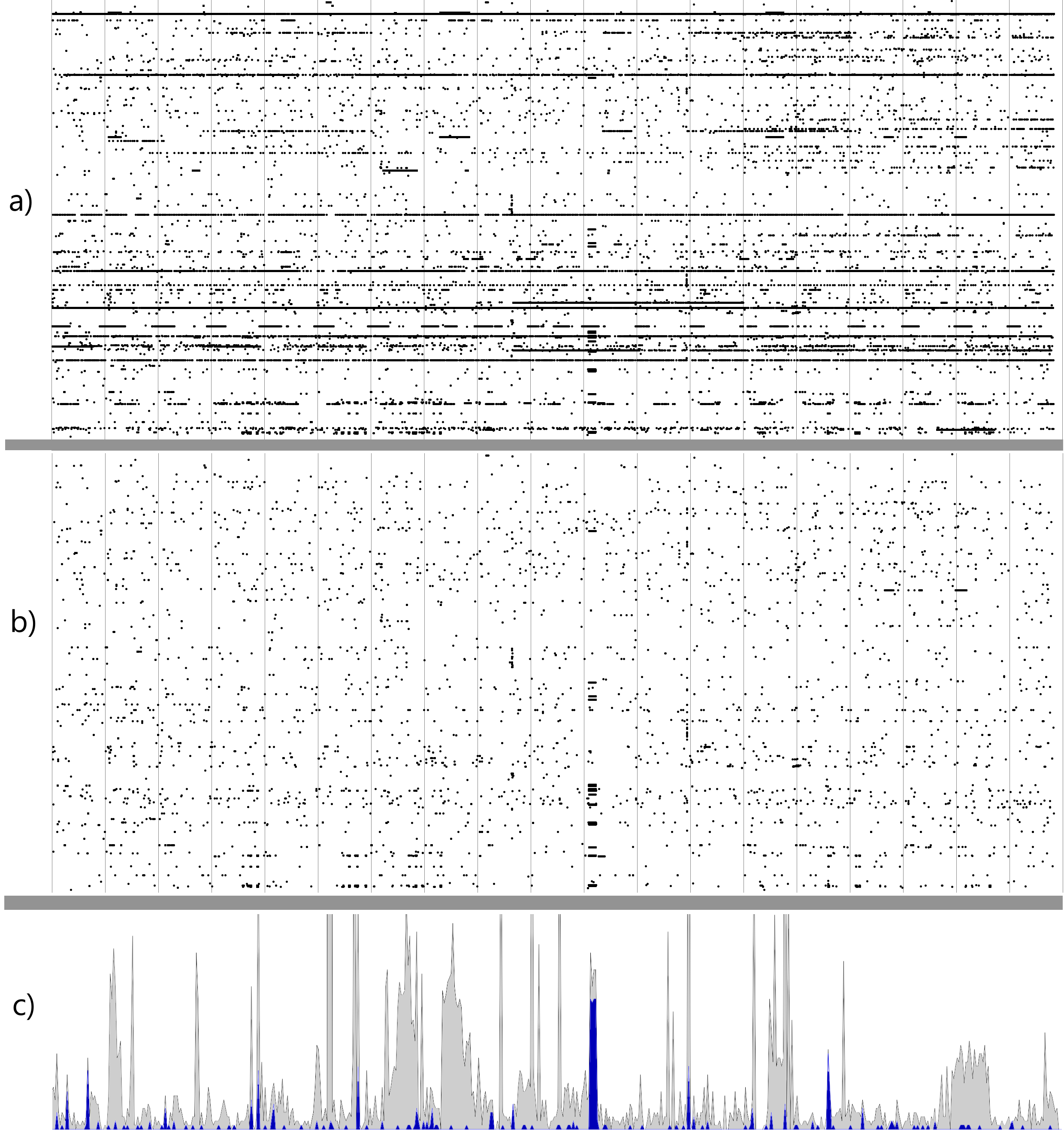}  
  \caption{Results of our technique are analyzed visually in access scatter plots. a) Incoming events show significant, periodic and mechanical access patterns consistent with AI crawlers. b) After hierarchical filtering using our metrics, these mechanical patterns are greatly reduced while still retaining the random, infrequent, short term access events more consistent with human behavior. c) Results show a 94\% reduction in estimated server workload, from gray to blue, measured as the total momentary server activity required to meet requests using a fixed response duration.}
  \label{fig_results}
\end{figure}

\section{Results}
\label{results}
Results reveal that mechanical access patterns are more thoroughly blocked with our method than with throttling alone. We valid this in several ways. For the case study of the Community Science Institute the total blocking actions would reduced the traffic by 94\%, Figure \ref{fig_results}b, which corresponds to the 80-95\% non-human access originally estimated. Visual analysis further confirms that mechanical patterns are eliminated whereas the more random, human-like patterns are retained. 

Incoming activity before blocking, Figure \ref{fig_results}a, shows clear indications of substantial AI crawling and bot activity. More subtly, in addition to continuous, repetitive patterns over many days (marker A), we also observe sustained daily partial activity over many days (marker F), activity with varying repetition rates (marker C) and irregular yet persistent activity (marker E). We also noticed distributed botnets using both class C subnets (marker G) and class B subnets (marker H).

Figure \ref{fig_results}b shows the results of applying the blocklist from our method to this log period. The majority of horizontal, repetitive patterns have been removed, while retaining the sporadic, momentary accesses typical of human activity. While some repetition is still noticeable the overall AI crawler and bot activity is greatly reduced, even those originating from class B and C subnets. 

The blocklists produced by Logrip were finally employed on two different real world servers. In both instances, improvements in the responsive of the servers was immediately observed. 

\begin{figure}[H]
  \includegraphics[width=\textwidth]{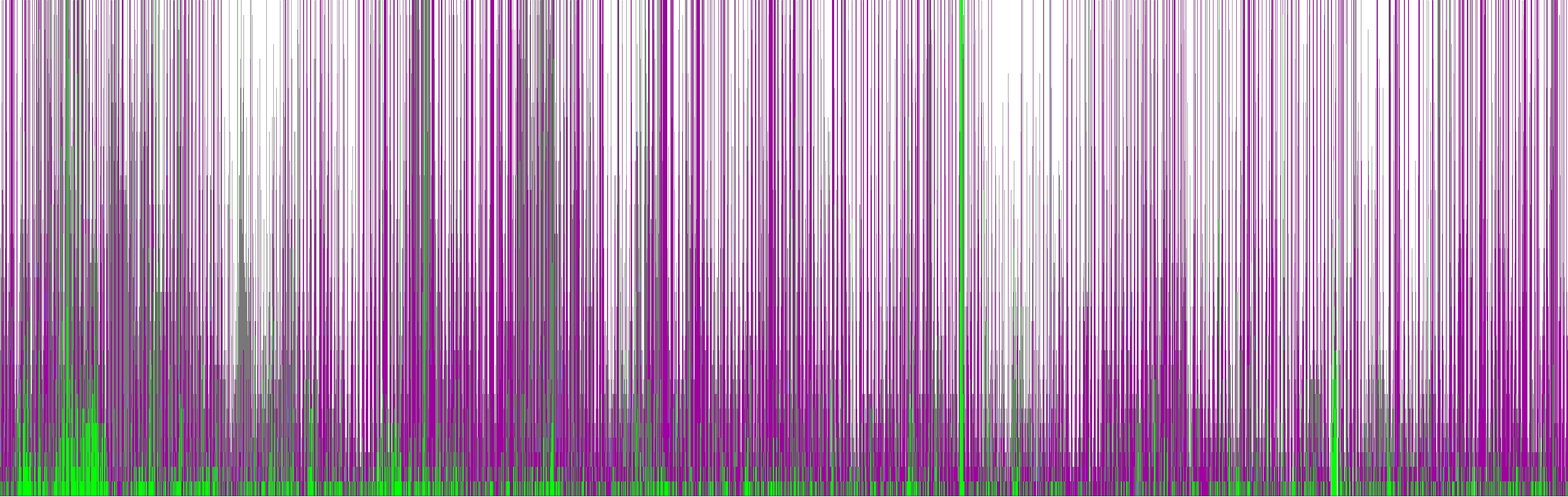}  
  \caption{Visualization of estimated server workload, with time on the X-axis over 20 days, and average requests per minute on the Y-axis. Grey depicts IP level blocked traffic and purple represents C-subnet level blocked data center traffic. Final, non-blocked traffic is bright green. These estimates assume a fixed server response duration, see Section 6.1. }
  \label{fig_workload}
\end{figure}

\subsection{Workload}
Server workload can be estimated by computing the momentary impact at time $t$ for all currently active access requests sustained for a fixed duration $ds$. While each real request will vary in response time, we can provide a rough estimate by assuming requests take constant time on average. The TCP inactivity timeout setting, typically 60 seconds, provides a good baseline for $ds$ since clients will remain connected even if inactive over this range until the TCP connect timeout is reached. Workload estimates thus include handshaking and not just server response in units of active average requests per minute, as shown in Figure \ref{fig_workload}.

Table \ref{tab:workload} shows the total integrated load over the 20 day duration, separated by algorithm filter stage. While the value itself is somewhat arbitrary due to the non-realistic fixed duration to estimate response times, the relative reduction in load - indicated as stage \% - is a good indicator of the amount of activity reduced by blocking of that metric. 

Throttling reduces the workload by 33\%, yet this is partly due to the fact that a larger number of rapid requests have been eliminated over a short time. Throttling effectively removes single, very bad actors, but does not address the total volume of automated crawling. Consecutive days, daily ranges, and daily maximums each reduced the workload by 9\%, 9\% and 3\% for this example data set. These metrics each contribute to removing crawlers and bots that exhibit different types of access.

Applying the same metrics to class C and class B subnets further reduces the workload by 14\% and 26\% respectively, accounting for a full 40\% reduction when using hierarchical hashing and blocking. Collectively, where single IP metrics resulted in a total 54\% reduction, including the subnet blocklists resulted in a cumulative 94\% reduction in workload. This corresponds very well our empirical estimate that 80-95\% of the observed traffic was likely AI crawlers and bots.

\begin{table}
  \begin{center}
  \label{tab:workload}
  \begin{tabular}{lrrr}
  \toprule
  Filter & Workload & Stage & Cumulative  \\
  & (ave. requests per min) & Reduction \% & Reduction \% \\
\midrule
None & 10.7 & & \\
Throttling & 7.1 & 33\% & 33\% \\
Consecutive & 6.2 & 9\% & 42\% \\
Daily range & 5.2 & 9\% & 51\% \\
Daily max & 4.9 & 3\% & 54\% \\
C Subnet & 3.4 & 14\% & 68\% \\
B Subnet & 0.6 & 26\% & 94\% \\
  \bottomrule
  \end{tabular}
  \end{center}
  \vspace{1em}
  \caption{Request handling workload reductions after each filtering stage of the algorithm. Stage percentage is the relative reduction for that single stage, while cumulative percentage shows the total reduction as all stage are included. The final cumulative reduction, 94\%, is the estimated drop in server workload after applying the final blocklist.}
\end{table}

\begin{figure}
  \begin{center}
  \includegraphics[width=0.9\textwidth]{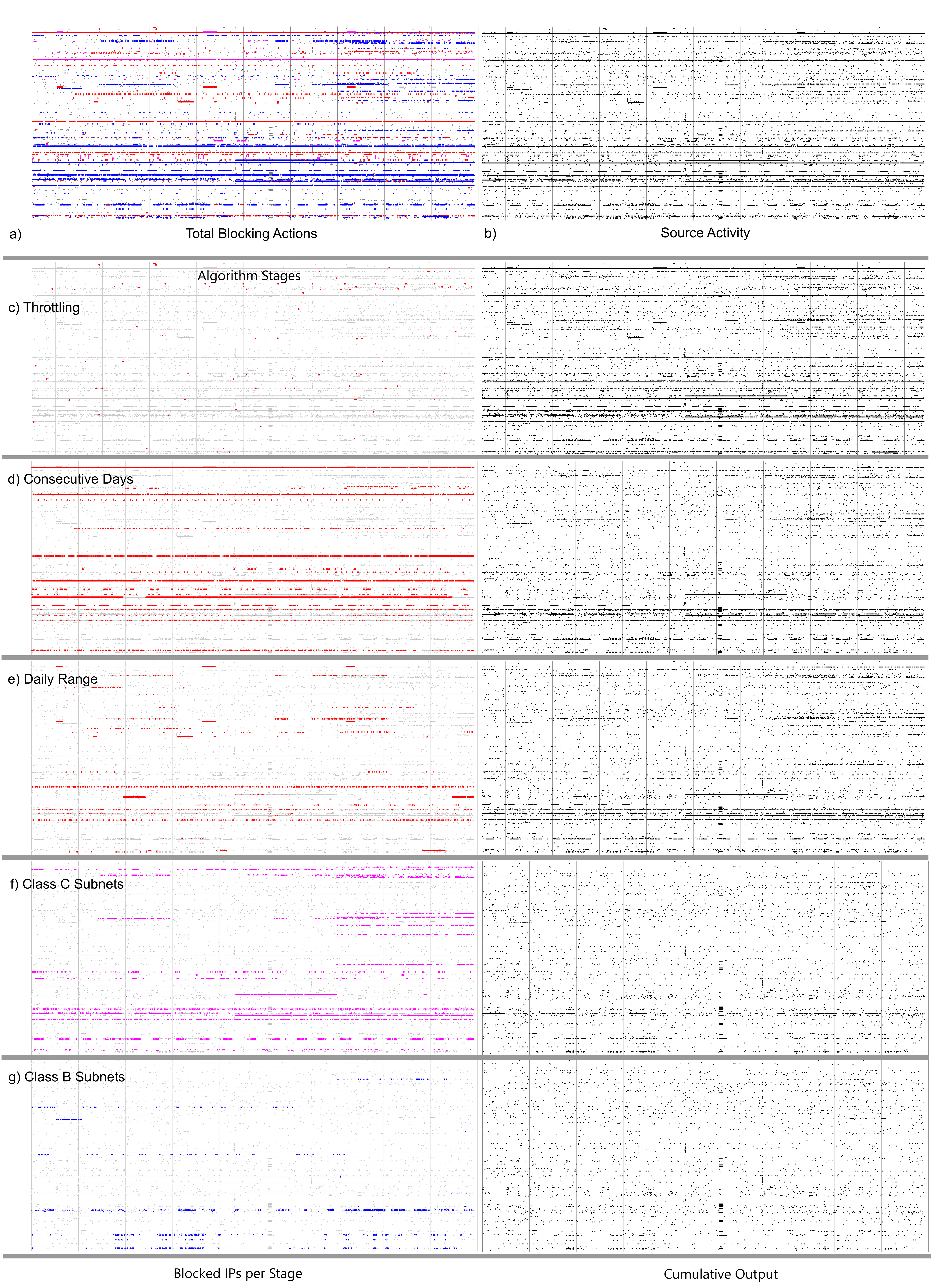}  
  \end{center}
  \caption{Data visualization of our algorithm as it proceeds through each filtering stage. a) The total blocking actions, top left, taken over the b) source activity from access logs, top right, are broken down into the filtering stages of c) throttling metrics, d) consecutive day limits, and e) daily range limits. These metrics are then applied hierarchically to remove f) class-C subnets, and g) class-B subnets that also show mechanical behavior. The effects of cumulative filtering of each stage are shown at right, with final output at the bottom right. (Zoom in for details)}
  \label{fig_stages}
\end{figure}

\subsection{Analysis of Algorithm Stages}
A more detailed analysis was performed by separating our algorithm into stages and visualizing blocking activity and cumulative filtering in the output, Figure \ref{fig_stages}. The top left shows total blocking activity, Fig \ref{fig_stages}a, which is broken into multiple stages, Fig \ref{fig_stages}c-g. The top right shows the original source activity, while the bottom right shows the cumulative output after each stage.

Results from the first stage, Figure \ref{fig_stages}c, confirm that throttling is ineffective at stopping web crawlers and botnets (blocked left, output right). While a few IPs had access rates higher than 20 pages per minute these are generally short lived. The majority of repetitive access is periodic yet well above rate limits. 

The next stage, Figure \ref{fig_stages}d, with high daily ranges over multiple consecutive days, is able to remove a significant amount of AI traffic. This metric was able to block these IPs even though access was below rate limits yet persistent over many days. Despite this, Figure \ref{fig_stages}d (right) shows that some mechanical patterns remains at low IP ranges and during intense portions on just a few days.

Daily range metrics, Figure \ref{fig_stages}e, not tied to consecutive days, are able to remove repetitive traffic that is established well above rate limits yet active for more than 6 hours per day. This handles web crawlers that may touch a site for a sustained portion of one day and then vanish. 

Despite these new single IP metrics, significant activity can still be observed, Fig \ref{fig_stages}e (right). By applying the same metrics to the class C subnet hash, we are able to block the majority of the remaining data center based AI crawlers, Figure \ref{fig_stages}f (left, pink). Applying the metrics to class B subnets, Figure \ref{fig_stages}g, further reduces any distributed, repetitive traffic over larger IP ranges.

\section{Discussion}
This work focuses on protecting single servers and small organizations from large, distributed AI crawlers operating from data centers. The exponential growth of AI-powered web crawlers underscores the urgency of our approach. Even well-behaved crawlers, when operating at scale, can disrupt small organizations and greatly impact server performance. Our methodology aims to empower organizations with the ability to detect, analyze, and regulate automated traffic in a manner that traditional security solutions do not address.

Logrip is presented as an open source, lightweight, command line tool for analyzing historic access logs and generating blocklists based on policy parameters. Our tool quickly identifies mechanical access patterns, even across data center subnets, and provides visual feedback of the pre-filtered and post-filtered blocking actions taken. Estimates of server load reductions are also output, which show that in many real world cases traffic reductions of 90-95\% are possible. 

Policy decisions are central to automated blocking activity. We develop a technique that allows controlled adjustment of policy parameters with visualization of their effects on blocking actions. In our case study the Community Science Institute makes its data publicly available, yet requires automated blocking of AI crawlers in order to defend against excessive traffic that degrades server performance for real human users. For many institutions some form of protection from persistent non-human traffic has become necessary.

\section{Limitations \& Future Work}
A significant limitation of the current study is the lack of verified ground truth data. Since our work relies on a real world case study, a baseline distinction between humans and machines was not available. In the future we would like to perform online and simulated experiments to determine the theoretical effectiveness of our techniques. 

A known issue with IP blocklists is their inability to prevent dynamic IP attacks such as Distributed Denial of Service (DDoS) attacks. Our tool also reveals this issue. In region J of Figure 1, a DDoS attack can be observed utilizing random, dynamic IPs across the entire IP spectrum. Since these IPs are dynamic, and the duration is short, our tool cannot block them and other methods are required for such attacks.

Our technique relies on access patterns alone, which we view as both a feature and a limitation. We intentionally disregard the nature of requests since we observed in our case study that benign web crawlers - accessing valid pages, without injection, and observing rate limits - may still overwhelm a server due to sheer volume. Therefore we seek to block mechanical patterns regardless of intent. Nonetheless, further investigation of the actual requests could enhance our technique.

The study of human versus machine access patterns presents a new perspective on Internet security. While the security field has largely focused on overtly malicious activity we believe that the massive increase in "benign" web crawlers seeking AI training data is not entirely altruistic. We hope this work encourages new research that considers the study of human versus machine activity as interesting in and of itself. 

\section{Funding}
This research did not receive any specific grant from funding agencies in the public, commercial, or not-for-profit sectors.


\appendix
\label{apxA}

\clearpage

\bibliographystyle{unsrtnat}
\bibliography{ms}  



\end{document}